\documentclass[9pt,twocolumn]{article}

\usepackage[hidelinks,unicode=true]{hyperref}
\hypersetup{
colorlinks=true,
linkcolor=blue,
urlcolor=blue,
citecolor=blue,
pdfhighlight=/N}
\usepackage{newtxtext,newtxmath}
\usepackage{xr}
\usepackage{authblk}
\usepackage[numbers]{natbib}
\usepackage{xcolor}
\providecommand{\keywords}[1]
{
  \small	
  \textbf{\textit{Keywords---}} #1
}
\usepackage{graphicx}
\usepackage{dcolumn}
\usepackage{bm}
\usepackage{amsmath}

\graphicspath{{figs}}

\usepackage{soul}

\begin{document}

\title{Data-Driven Stochastic Modeling of Schooling Fish: From Collective Dynamics to Individual Fluctuations}

\author[a]{Elena G. de Lamo}
\author[b,c]{M. Carmen Miguel}
\author[a]{Romualdo Pastor-Satorras}

\affil[a]{Departament de Física, Universitat Politècnica de Catalunya, Campus
  Nord B4, 08034 Barcelona, Spain}
\affil[b]{Departament de Física de la Matèria Condensada, Universitat de
  Barcelona, Martí i Franquès 1, 08028 Barcelona, Spain}
\affil[c]{Institute of Complex Systems (UBICS), Universitat de Barcelona,
  08028 Barcelona, Spain}
\renewcommand\Affilfont{\itshape\small}

\twocolumn[
  \begin{@twocolumnfalse}
\maketitle

\begin{abstract}
Collective motion in animal groups emerges from the interplay between individual variability and social coordination, yet connecting these scales quantitatively has remained a major challenge.
Using high-resolution trajectories of schooling fish, we infer a data-driven stochastic framework that reproduces with remarkable accuracy the behavior of real fish schools. We decompose motion into two coupled components: the dynamics of the school’s center of mass (or centroid), modeled as an active Brownian particle confined by the tank, and individual motions relative to that center, described by stochastic equations with data-inferred mean-field potentials and multiplicative noise. Simulations of these equations produce synthetic schools that quantitatively match real ones across multiple observables, including burst-and-coast dynamics, polarization, and spatial cohesion.
This minimal, predictive framework bridges experiment and theory, showing that the collective dynamics of animal groups can be faithfully reconstructed from first principles directly from data.

\end{abstract}
 \end{@twocolumnfalse}
]

Fish schools exhibit a remarkable capacity to move as cohesive, coordinated groups, performing synchronized turns, accelerations, and scattering maneuvers seemingly without a leader or any centralized control~\cite{sumpterCollectiveAnimalBehavior2010}. Deciphering how such order arises from local interactions among inherently noisy, autonomous individuals remains a central challenge in the study of collective behavior~\cite{vicsek_collective_2012}.
Beyond its ecological importance in facilitating predator avoidance, migration, and foraging, collective motion has proven to be a source of cross-fertilization in fields such as bioinspired algorithms for swarm robotics, distributed decision-making and related fields~\cite{couzinEffectiveLeadershipDecisionmaking2005,BrambillaSwarmRobotics2013,vicsekCollectiveMotion2012,puySelectiveSocialInteractions2024}.

Here we show that a minimal, data-driven stochastic model can quantitatively reproduce the collective dynamics of real fish schools.
Starting from empirical trajectories, we infer effective mean-field potentials and state-dependent (multiplicative) noise terms that capture both the group’s global movement and the local fluctuations of individual fish.
Simulations of the inferred equations produce synthetic schools that match real ones across multiple observables---including burst-and-coast dynamics, long-term displacement patterns, polarization, and spatial cohesion---demonstrating that the essential ingredients of collective motion can be recovered directly from data without ad hoc behavioral rules or tuned parameters.

Models of collective motion have a long lineage, from agent-based alignment rules exemplified by the Vicsek model~\cite{vicsekCollectiveMotion2012} to social‑force formalisms and continuum hydrodynamic descriptions. These frameworks have successfully captured many qualitative phenomena—group cohesion, order–disorder transitions, spontaneous collective turns, and related instabilities—but they typically rest on phenomenological interaction rules whose functional forms and parameters are chosen a priori or tuned to reproduce particular observations. That reliance on assumed rules and fitted parameters constrains their predictive scope and complicates mechanistic interpretation. An alternative strategy is to infer directly from experimental trajectory data the effective stochastic laws that govern motion, thereby extracting the mean forces and fluctuation statistics that drive collective behavior~\cite{Herasdeepattentionnetworks2019,kelley2013}.  Data‑driven inference thus provides a principled route to connect microscopic variability and interaction-induced correlations with macroscopic order, forging a tighter link between experimental observation and the statistical‑physics description of moving animals.

Our framework implements this strategy by describing the motion of a fish school as a stochastic process evolving on two coupled levels of organization.
At the collective level, the school’s center of mass (CM) behaves as a finite-size active particle undergoing Brownian-like motion under confinement imposed by the tank boundaries.
At the individual level, each fish moves stochastically relative to the CM, governed by an effective confining potential and state-dependent (multiplicative) noise. Individuals move freely within the group but are gently pulled back by a near‑constant force when they stray beyond its boundary, plausibly because vision—--dominant at these scales—--maintains cohesion while visual contact persists regardless of distance. The deterministic (drift) and stochastic (diffusion) components at both scales are inferred from experimental trajectories using a nonparametric kernel regression approach~\cite{lamouroux_kernel-based_2009}. This decomposition provides a compact, interpretable description of the group's dynamics that links individual fluctuations to emergent collective behavior without assuming prescribed interaction rules.

This two-level description captures a fundamental feature of animal groups: the coexistence of order and variability.
The CM motion encodes the coherent, correlated displacements of the entire group, while the relative fluctuations reveal how individuals explore the neighborhood of the collective state.
Within this framework, multiplicative noise naturally emerges as a key organizing element.
Unlike additive noise, whose strength is independent of the system’s state, multiplicative noise depends on the instantaneous configuration of the group and thus modulates the intensity of fluctuations.
This mechanism can generate rich dynamical phenomena—such as intermittent turning cascades, collective transitions, and bursts of coordinated activity—that are widely observed in real fish schools~\cite{puy_signatures_2024,mugica_scale-free_2022}.
Similar noise-induced effects play central roles in other complex systems, from neural synchronization to population dynamics~\cite{horsthemke2006noise,bauermannMultiplicative2019}, underscoring the broad relevance of this stochastic framework.

By successfully reproducing the statistical and dynamical properties of real fish schools from first principles, our results show that collective motion can be both understood and quantitatively predicted by a data-driven stochastic model grounded in statistical physics.
This approach closes the loop between empirical data, theory, and simulation: empirical  trajectories provide effective drift and diffusion functions, which in turn generate synthetic schools whose emergent behavior reproduces experiments across scales. More broadly, our findings deepen the theoretical understanding of self-organization in biological collectives and provide a principled basis for extending these methods to other systems of interacting agents.

\section*{Modeling framework}

To analyze the movement of schooling fish within an experimental tank, we decompose each fish’s laboratory-frame trajectory, denoted as $\vec{r}_i^\mathrm{lab}(t)$, into a collective and a relative component:
\begin{equation}
  \label{eq:break_path}
  \vec{r}_i^\mathrm{lab}(t)  = \vec{R}(t) + \vec{r}_i(t).
\end{equation}
Here $\vec{R}(t) = \sum_{i=1}^N\vec{r}_i^\mathrm{lab}(t) /N$ represents the position of the group's CM (considering all fish with mass $m=1$), with $N$ the number of fish, and describes the school’s collective position at any time $t$, while $\vec{r}_i(t) = \vec{r}_i^\mathrm{lab}(t) - \vec{R}(t)$, denotes the motion of fish $i$ relative to the CM. This separation isolates collective dynamics from individual fluctuations and follows previous work on correlated motion in animal groups~\cite{cavagna_diffusion_2013,murakami_inherent_2015}.
To empirically investigate both the behavior of the CM trajectory and the internal dynamics relative to it, we analyzed the motion of freely swimming individuals in schools (of size $N=40$, $50$, and $60$)  of black neon tetra (\textit{Hyphessobrycon herbertaxelrodi})  a species known for its strong schooling behavior and responsiveness to environmental cues.  For full experimental details, see Methods. 
Based on these  experiments, we propose two sets of stochastic differential equations~\cite{oksendalStochastic2014,mendezStochasticFoundationsMovement2014} whose functional forms and parameters are inferred from the empirical data.

\subsection*{Center of mass dynamics}

\begin{figure}[t]
  \centering
  \includegraphics[width=\columnwidth]{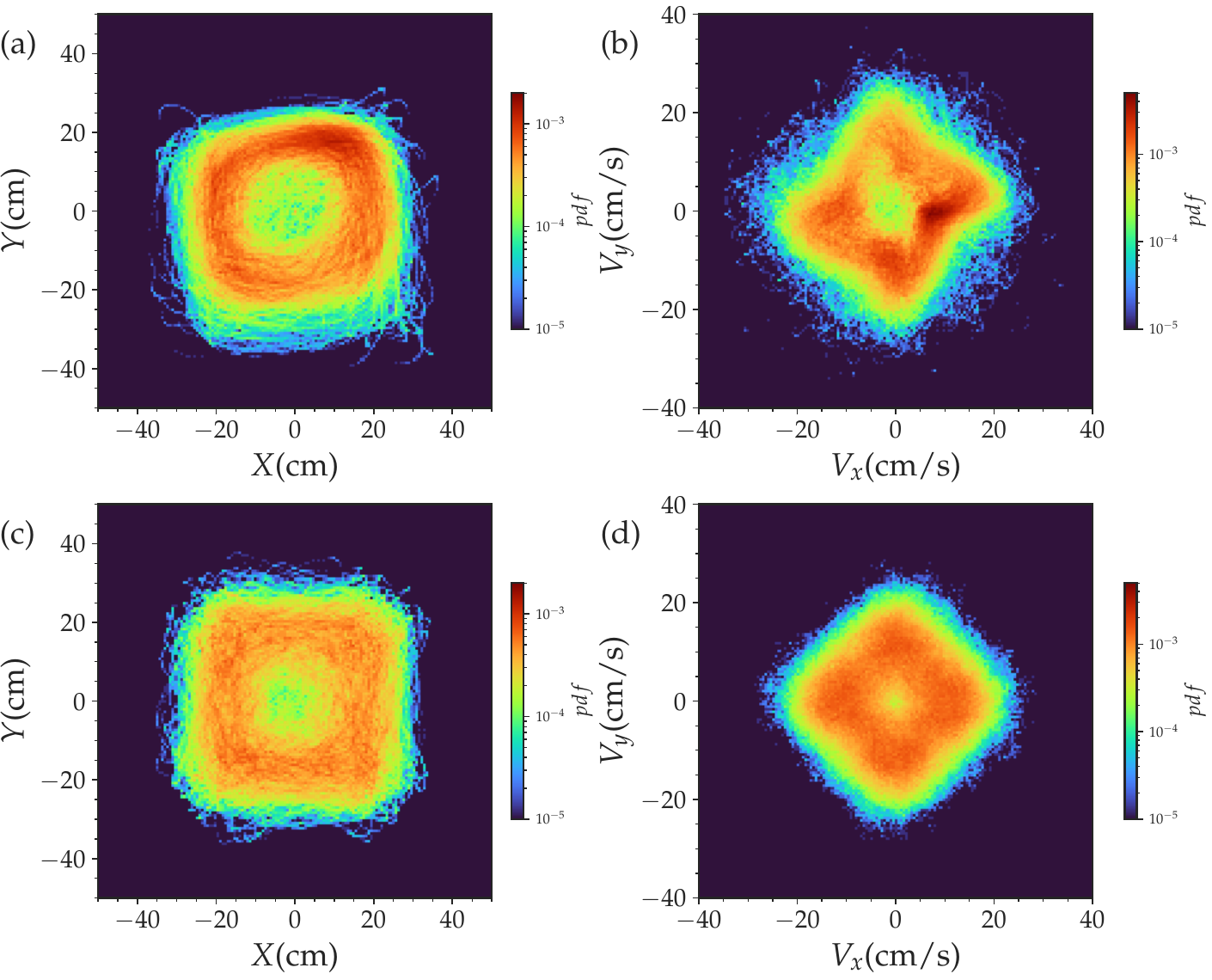}%
  \caption{Probability density functions (PDFs) of the school centroid (CM) position (a) and velocity (b) from empirical data, aggregated over three experimental realizations ($N=60$). Panels (c) and (d) show the corresponding PDFs obtained from numerical simulations. Results for other group sizes are presented in Supplemental Figure~\protect SF1. \label{fig:1} }
\end{figure}

In the upper panels of Fig.~\ref{fig:1}, we present color density plots of the probability density functions (PDFs) for the CM’s position and velocity.
The figure reveals that the CM trajectory exhibits persistent motion, constrained by the boundaries of the experimental tank, with a tendendy to slide along the walls, a behavior consistent with previous observations in real fish~\cite{calovi_disentangling_2018, Lecheval2023}. 
This wall-following tendency reflects the interplay between the self-propelled dynamics of the group and the confining geometry of the environment. 
On the other hand, the PDF of the CM velocity exhibits a characteristic crater-shaped profile, typical of self-propelled active particles~\cite{romanczukActiveBrownianParticles2012}. Additionally, the velocity components along the principal axes can be accurately modeled using a Gaussian mixture, a feature also observed in other systems of self-propelled particles~\cite{erdmann2000brownian-fc3}, see Supplementary Figure~SF3.

These observations indicate that the movement of the school’s CM can be understood within the framework of active matter physics, influenced by confinement forces arising from the boundary effects of the tank walls. 
We thus choose to parsimoniously model the CM dynamics using a set of stochastic differential equations that describe an active Brownian pseudo-particle with a characteristic size corresponding to the average radius of gyration of the school, $R_g$,
of the form
\begin{equation}
    \frac{d\vec{R}}{dt}  = \vec{V}, \qquad
    \frac{d\vec{V}}{dt} = \vec{F}(\vec{R}) - \gamma(V) \vec{V} + \sqrt{2 D}
    \xi(t).
  \label{eq:main_1}
\end{equation}
The assumption of a finite size naturally arises from the fact that the CM of the school cannot approach the walls as closely as any individual fish. We consider all fish  have the same mass, an the total mass has been absorbed into the coefficients on the right-hand side of Eq.~\eqref{eq:main_1}.
This pseudo-particle is subjected to a confining force—--originating from the boundaries of the tank—--as well as an active friction term and a noise term characteristic of self-propelled systems~\cite{romanczukActiveBrownianParticles2012}.
For the active friction, we choose a Schienbein-Gruler-type form~\cite{Schienbein1993} $\gamma(V)\vec{V}  = -\gamma_0 (V - V_0) \frac{\vec{V}}{V}$, with $V$ the CM's velocity modulus.
For the confining force, we adopt the non-potential hard-wall model proposed in Ref.~\cite{calovi_disentangling_2018} to account for the tendency of fish to move parallel to walls:
\begin{equation}
  \label{eq:main_2}
  \vec{F}(\vec{R}) = U_0 \sum_{w \in \mathrm{Walls}} \exp \left\{\frac{R_g -
      d_{w}}{\ell}\right\}g(\theta_w) \hat{n}_w,
\end{equation}
where the function  $g(\theta_w) = 1 + \lambda \cos(2 \theta_w)$
modulates the force, making it weaker when the pseudo-particle moves parallel to the wall and stronger if it moves toward or away from it. The parameters in the model are estimated from the experimental velocity profile and by minimizing the Kullback-Leibler divergence  between the steady-state PDFs of position and velocity components generated by simulations and those extracted from the experimental data~\cite{burnham2002model}. For details on the model definition  and parameter fitting procedure, see Supplementary Material~SM-I. The optimal values obtained are reported in the Supplementary Table~ST1.

\subsection*{Relative individual movement in the CM frame}

\begin{figure}[t]
  \centering
  \includegraphics[width=\columnwidth]{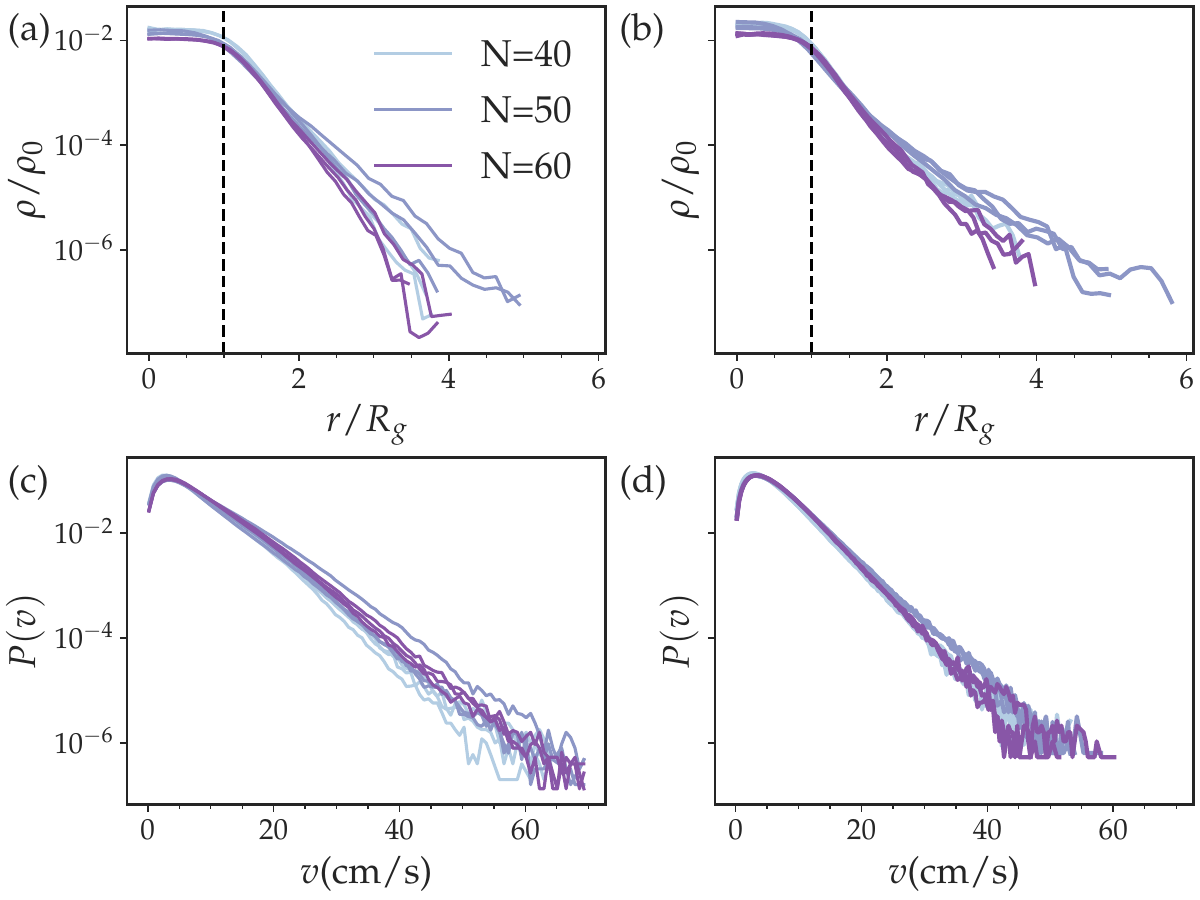}%
  \caption{
   Radial density of fish in the CM frame for (a) experimental data and (b) model simulations (b). Densities are normalized by each realization’s radius of gyration $R_g$ and by $\rho_0$ so curves collapse. Dashed lines mark $r=R_g$. (c) Speed distribution in the CM frame from experimental data. (d) Corresponding speed distribution from the model.}
  \label{fig:2} 
\end{figure}

In Supplementary Fig.~SF5 
we present color density plots of the PDFs of fish positions and velocities in the CM frame. 
Both distributions display clear rotational symmetry. From the position PDF we compute the radial density, $\rho(r)$, shown in Fig.~\ref{fig:2}(b). Across different school sizes this quantity displays a universal scaling behavior expressed as $\rho(r) = \rho_0 F(r/R_g)$, where $\rho_0 = N / (\pi R_g^2)$ is the average density for each data set. The scaling function $F(z)$ is approximately constant for small $z$ and decays roughly exponentially for large arguments.  

The velocity PDFs along the Cartesian axes, shown in Fig.~\ref{fig:2}(d), exhibit pronounced exponential tails rather than Gaussian ones, suggesting the presence of multiplicative (state-dependent) noise. Exponentially-tailed speed distributions have been reported in other collective systems, for example large midge swarms~\cite{kelley2013}.  In this context, the impact of multiplicative noise has been analyzed previously \cite{reynolds2016}. Likewise, studies of migrating social amoebae found non-Gaussian velocity components and modeled them using Langevin dynamics with multiplicative noise~\cite{bodeker_quantitative_2010}, a framework that naturally captures the observed deviations from Gaussianity.

To model fish behavior in the center of mass (CM) frame, we employ a system of stochastic equations in polar coordinates. This system describes the velocity modulus or speed $v$, the orientation of the velocity vector $\theta$, and the turning rate $\omega$, which is related to the angular velocity of the velocity vector~\cite{zienkiewicz_leadership_2015, gautraisDecipheringInteractionsMoving2012}. Supplemental Fig.~SF6  illustrates the PDF of the turning rate $\omega$, which also exhibits a non-Gaussian form. 
In Supplemental Fig.~SF7, we examine the time correlations and cross-correlations between $v$ and $\omega$. This figure reveals that the empirically measured speed and turning rate show weak cross-correlations between different fish, compared to the strong auto-correlation observed for the same fish.
This observation suggests that the behavior of a fish interacting with other members of the school can be effectively represented as that of an {\em isolated} individual subject to a potential that, in a mean-field approximation, captures the influence of the group. While this approximation is a priory quite stringent, we will demonstrate that it allows us to reproduce many features observed in empirical schools.

We model the movement of individual fish as independent in the CM reference frame, each subject to an effective central confining potential $U_\mathrm{eff}(r)$ that prevents drift away from the CM. The stochastic equations in polar coordinates, representing the relative motion in the CM, are chosen as
\begin{equation}
  \begin{split}
    \dot{x} &= v \cos \theta, \qquad \dot{y} = v \sin \theta,\\
    \dot{v} &= F_\mathrm{eff}^v+ f_v(v) + g_v(v) \; \xi_v(t),\\
    \dot{\theta} &= \omega +  F_\mathrm{eff}^\theta, \\
    \dot{\omega} &= f_\omega(\omega) + g_\omega(\omega) \; \xi_\omega(t).
  \end{split}
  \label{eq:4}
\end{equation}
The terms $F_\mathrm{eff}^v$ and $F_\mathrm{eff}^\theta$ represent the forces induced by the confining potential when expressed in polar coordinates, while the terms $f_y(y)$ and $g_y(y)$ ($y = v, \omega$) represent possible drift and diffusion components induced by interactions with the rest of fish.

To estimate the form of this effective confining potential, we draw on the analogy with equilibrium Brownian motion in a potential $U$ with constant friction~\cite{romanczukActiveBrownianParticles2012}, for which the radial distribution is proportional to $e^{-U(r)/A}$ with $A$ constant. Further support for this assumption is provided by an analysis of the radial flux of fish entering and leaving a circular region of radius $R_g$, see Supplementary Material~SM-II. Thus, assuming this relation holds in our context, we define the  confining effective potential as $\exp \left\{ - U_\mathrm{eff}(r)/A\right\} = \rho(r)$.
The empirical $\rho(r)$ (see Fig.~\ref{fig:2}(b)) is nearly flat for $r < R_g$ and decays exponentially for $r>R_g$. 
Therefore, we choose to fit it with the continuous function
\begin{equation}
  \label{eq:1}
  \exp \left\{ - U_\mathrm{eff}(r)/A\right\} 
  \equiv \frac{a}{b +
    e^{\alpha r}},
\end{equation}
where the parameters $a$, $b$, and $\alpha$ are obtained from a non-linear curve fitting. From this central potential, the effective force takes the form 
\begin{equation}
  \label{eq:2}
  \vec{f}_\mathrm{eff}(r) = -\frac{\partial U_\mathrm{eff}(r)}{\partial r}
  \hat{r} 
  \equiv -A  f_\mathrm{eff}(r) \; \hat{r} ,
\end{equation}
where $\hat{r} = \vec{r}/r$, $f_\mathrm{eff}(r) = \frac{\alpha}{b e^{-\alpha r} + 1}$, and the effective terms $F_\mathrm{eff}^v$ and $F_\mathrm{eff}^\theta$ in~\eqref{eq:4} can be written as
\begin{equation}
  \label{eq:3}
  F_\mathrm{eff}^v =  A   (-\hat{v} \cdot \hat{r}) \; f_\mathrm{eff}(r), \quad
  F_\mathrm{eff}^\theta =  A   \frac{1}{v} (-\hat{v} \times \hat{r}) \;
  f_\mathrm{eff}(r).
\end{equation}
Inside the group, individuals move unhindered, but once they pass the boundary they are drawn back by an approximately constant restoring force, likely because vision—dominant at these scales—helps preserve cohesion regardless of distance. The larger $R_g$ the weaker the restoring force.

Finally, the terms $f_\alpha$ and $g_\alpha$ (for $\alpha = v, \omega$), which represent drift and diffusion contributions arising from additional fish interactions~\cite{romanczukBrownianMotionActive2011}, are estimated empirically from data segments where the confining potential is negligible. Concretely, we use trajectories with $r < R_g$, for which the trapping potential is effectively flat and its force can be ignored, to extract the interaction-induced drift $f_\alpha$ and the state-dependent diffusion $g_\alpha$. 
Assuming that the stochastic equations for $v$ and $\omega$ stem from underlying Fokker-Planck equations, we  apply a kernel-based regression method~\cite{lamouroux_kernel-based_2009} to extract the first two moments of the Kramers-Moyal series, corresponding to the drift and diffusion coefficients. The resulting drift and diffusion coefficients, in the freely swimming regime, can be fitted to the functional forms: 
\begin{eqnarray}
    f_v(v) &=& \frac{\gamma_1}{v} - \gamma_2 v, \quad   g_v(v) = \delta_1 v + \delta_2, \\
    f_\omega(\omega) &=& - \kappa \omega, \quad   g_\omega(\omega) = n_1 |\omega| + n_2,
\end{eqnarray}
where $\gamma_i$, $\delta_i$, $\kappa$, and $n_i$ are empirically fitted constants. The specific parameter values for different experimental realizations are listed in  Supplemental Table~ST3.  See Supplementary Material~SM-II for further details on the model definition and the parameter fitting procedure.

\section*{Results}

\subsection*{Numerical check}

To check the models' dynamical accuracy at both collective and individual scales, we integrate the inferred stochastic equations numerically. For the collective-level set of equations \eqref{eq:main_1} we use the Euler-Maruyama scheme~\cite{toralStochasticNumericalMethods2014} with a timestep equal to the empirical frame interval $\Delta t= 0.02$s. For the local level equations~\eqref{eq:4}, which include multiplicative noise, we use the JitcSDE integration package~\cite{jitcxde}, since the plain Euler–Maruyama method is not appropriate for state-dependent noise~\cite{toralStochasticNumericalMethods2014}.

As we can see in Figs.~\ref{fig:1} and ~\ref{fig:2}, simulations accurately reproduce the empirical PDFs of position and velocity for the CM dynamics, as well as the radial density and velocity distributions of individual motion in the CM frame. Supplementary Fig.~SF13 also shows very good agreement between simulated and empirical turning-rate distributions in the CM frame (Supplementary Fig.~SF6).

\subsection*{Synthetic schools}

As we have seen, \eqref{eq:main_1} and~\eqref{eq:4} describe the movement of the school's CM and the relative motion of individual fish within the CM frame, respectively. Combining these equations, we can generate \textit{synthetic} schools of arbitrary size $N$ by: (i) simulating a single trajectory for the CM with~\eqref{eq:main_1}, and (ii) generating $N$ \textit{independent} trajectories in the CM frame using \eqref{eq:4}.  The trajectories of $N$ synthetic fish in the laboratory reference frame are then obtained  using~\eqref{eq:break_path}. 

The limit $N \to \infty$ provides analytic insight into very large synthetic schools. In this limit the radius of gyration effectively diverges and the confining potential $U_\mathrm{eff}$ becomes constant. Consequently the hard-wall force $\vec{F}(\vec{R})$ and the effective forces $F_\mathrm{eff}^v$ and $F_\mathrm{eff}^\theta$ vanish, reducing the dynamics to \eqref{eq:main_1} and~\eqref{eq:4} with those forces set to zero.

For the relative motion in the CM frame, the velocity and turning rate dynamics decouple. In the long-time (stationary) limit, the marginal stationary distribution for the speed $v$ takes the form  

\begin{equation}
    P_0(v) = \mathcal{N} \frac{
    v^{2 \gamma_1/\delta_2^2}}{(\delta_1 v +\delta_2)^{2(\beta_+ + 1)}} \exp\left\{2\delta_2\frac{ \beta_-}{(\delta_1 v +\delta_2)}\right\},
\label{eq:p_0_v_sup}
\end{equation}
where $\beta_\pm = (\gamma_1/\delta_2^2 \pm \gamma_2/\delta_1^2)$. Noticeably, an analogous expression for the speed PDF  was reported for migrating social amoebae~\cite{bodeker_quantitative_2010}.
Similarly, the stationary distribution for the turning rate $\omega$ is 
\begin{equation}
    P_0(\omega) = \frac{\mathcal{N} }{( n_1 |\omega| +n_2)^{2(1+\kappa/n_1^2)} }\exp\left\{-\frac{2 \kappa n_2}{n_1^2(n_2+ n_1 |\omega|)}\right\}.
    \label{eq:p_0_w_sup}
\end{equation}
See Supplementary Material~SM-III for mathematical details. 

The asymptotic forms for large $v$ and $\omega$ of the previous distributions can be easily recovered as
\begin{equation}
    P_0(v) \sim v^{-2(1 + \gamma_2 / \delta_1^2)}, \quad P_0(\omega) \sim |\omega|^{-2(1 + \kappa/n_1^2)}.
    \label{eq:asymptotes}
\end{equation}
From the data in Supplementary Table~ST3 we estimate the turning-rate tail exponent 
$P_0(\omega) \sim \omega^{-4.2}$, essentially invariant with group size. The speed tail exponent is much larger, ranging approximately from $P_0(v) \sim v^{-10}$ to $P_0(v) \sim v^{-13}$ across datasets.
Supplementary Fig~SF12  compares the analytical stationary PDFs from \eqref{eq:p_0_v_sup} and~\eqref{eq:p_0_w_sup} with empirical distributions of velocity $v$ and turning rate $\omega$, respectively, measured for trajectories confined to $r<R_g$. The analytical solutions closely match the empirical data; in particular, the turning-rate PDF exhibits the predicted power-law tail (\eqref{eq:asymptotes}). The velocity distribution appears to decay faster (effectively exponentially in the plotted range, see Fig.~\ref{fig:2}), consistent with the very large theoretical power-law exponent for $v$. 

\begin{figure}[t]
  \centering
  \includegraphics[width=\linewidth]{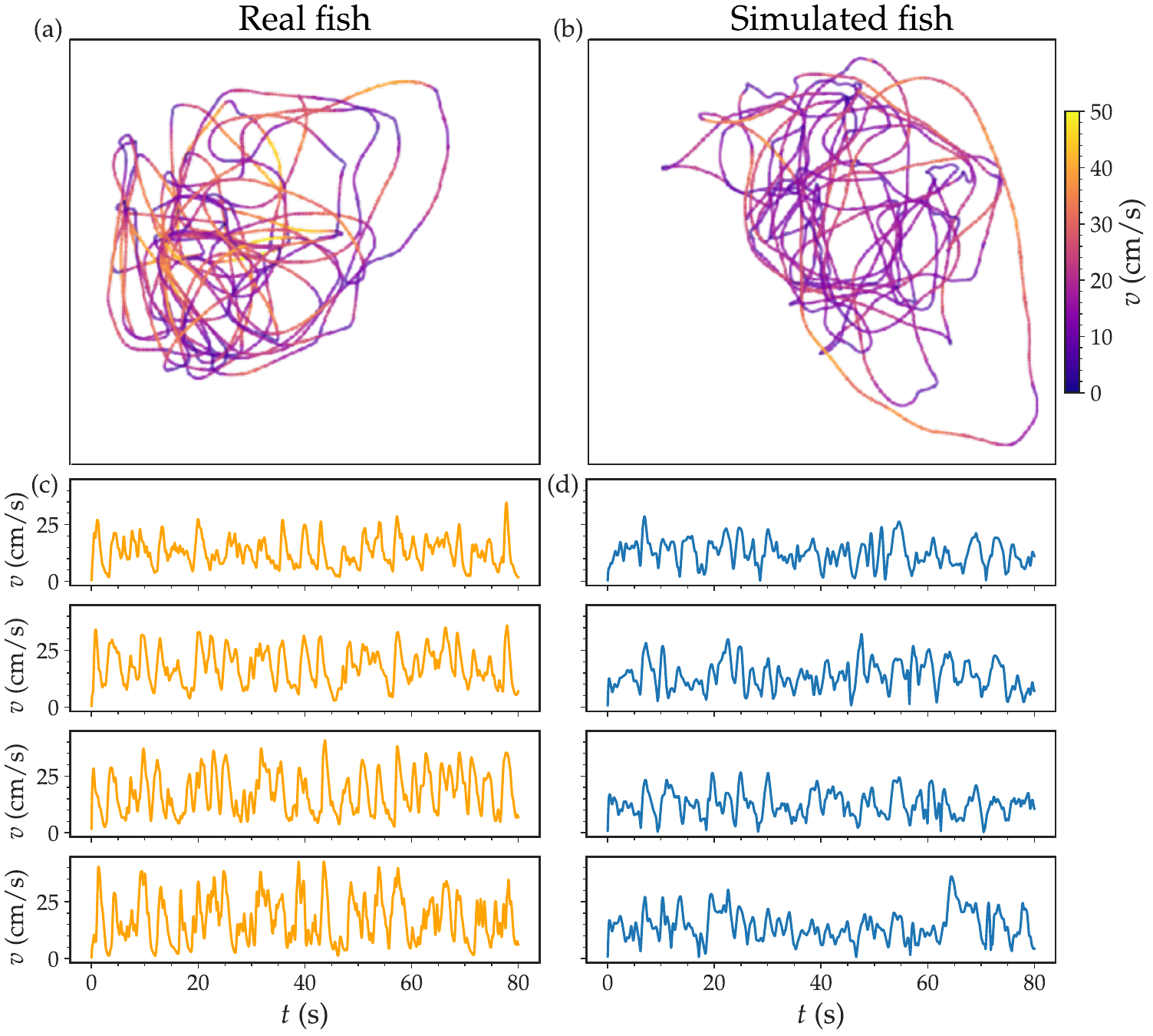}
  \caption{(a,b) Side-by-side comparison of a real fish trajectory and a simulated trajectory of equal duration (group size $N = 60$). (c,d) Temporal traces of speed for selected individuals from experiments (c) and simulations (d), illustrating synchronized burst‑and‑coast episodes. The same Gaussian smoothing applied to experimental data was also applied to simulation outputs for direct comparison.}
  \label{fig:trajectories}
\end{figure}

For finite $N$, numerical integration of the inferred stochastic differential equations yields synthetic schools whose individuals move in close agreement with real fish. As an example, Fig.~\ref{fig:trajectories}(a) presents a side-by-side comparison of an empirical trajectory and a model-generated trajectory; the similar motion patterns illustrate the model’s ability to reproduce naturalistic behavior.

Our model effectively captures the burst-and-coast swimming strategy observed in black neon tetra and many other fish species~\cite{Videler1982EnergeticAdvancesBurstandCoast}, characterized by alternating phases of rapid acceleration (bursts) and passive gliding (coasts). These speed oscillations often show partial synchronization: a burst by one fish can trigger similar accelerations in neighbors. This phenomenon has been studied in pairwise contexts~\cite{calovi_disentangling_2018}, in groups~\cite{harpaz_discrete_2017}, and even modeled using neural networks frameworks~\cite{gyllingbergUsingNeuronalModels2023}. As shown in  Fig.~\ref{fig:trajectories}(b), synchronized speed oscillations emerge naturally from our inferred stochastic dynamics.

Since the model describes a group of random walkers, it is natural to characterize their dynamics using the mean square displacement (MSD) as a function of time~\cite{gardiner_handbook_1985}. The MSD in the laboratory frame is defined as the ensemble average of the squared displacement of individual fish over time $t$, namely
\begin{equation}\label{eq:msd}
    \delta {\vec{r}^{\mathrm{lab}}}(t)^2  = \frac{1}{T-t} \frac{1}{N} \sum_{t_0=0}^{T-t-1} \sum_{i=1}^{N}\left[{\vec{r}_i}^{\mathrm{lab}}(t_0+t)-{\vec{r}_i}^{\mathrm{lab}}(t_0)\right]^2,
\end{equation}
where  ${\vec{r}_i}^{\mathrm{lab}}(t)$ denotes the position of individual $i$ at time $t$, and the average is taken  over the ensemble of individuals and over all time lags of duration $t$ within the observation window $[0,T]$. 
This definition of the MSD matches that used in~\cite{cavagna_diffusion_2013} for diffusion measurements in starling flocks, both in the laboratory and center-of-mass frames.

\begin{figure}[t]
    \centering
    \includegraphics[width=0.7\linewidth]{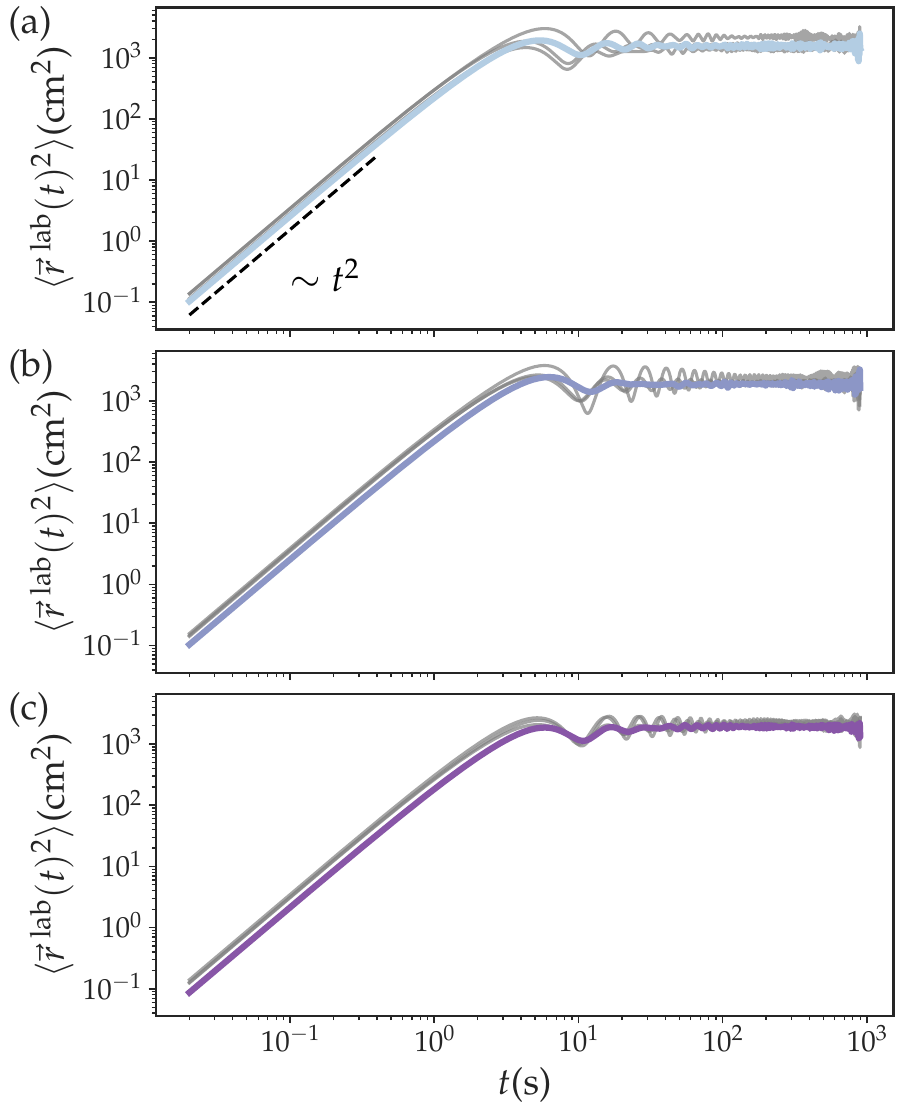}
    \caption{Mean square displacement (MSD) in the laboratory frame for (a) $N=40$, (b) $N=50$, and (c) $N=60$. Gray lines represent individual experimental realizations; colored lines depict the corresponding model predictions. MSD exhibits a short‑time ballistic regime and long‑time saturation due to tank confinement. Small oscillations in the plateau reflect collective circulation of the group's center of mass within the tank.}
    \label{fig:msd_lab}
\end{figure}

Using this MSD definition, we computed displacements from both experimental and simulated trajectories  (Fig.~\ref{fig:msd_lab}). In the laboratory frame the MSD shows oscillatory behavior, most visible in the plateau region; these reflect coherent, looping group motion constrained by the tank boundaries. The model reproduces these oscillations well, especially for the group of size $N=60$. For smaller groups, the experimental MSDs  display larger sample-to-sample variability due to fluctuations; while the model does not reproduce each realization exactly, it captures the intermediate trend across experiments.

As fish move within the group, individuals frequently cross in and out of the radius of gyration, $R_g$. To quantify these transitions, we compute the residence (or permanence) time, $\tau_{\mathrm{in}}$, defined as the duration a fish remains within $r < R_g$ after entering. Concretely, if a fish enters at time $t_0$ its residence time $\tau_{\mathrm{in}}$ is the time such that  $r_t < R_g$ for $t = t_0, \dots, t_0 + \tau_{\mathrm{in}} - 1$, and crosses the boundary ($r_T > R_g$) at time $T = t_0 + \tau_{\mathrm{in}}$. 

\begin{figure}[t]
    \centering
    \includegraphics[width=\linewidth]{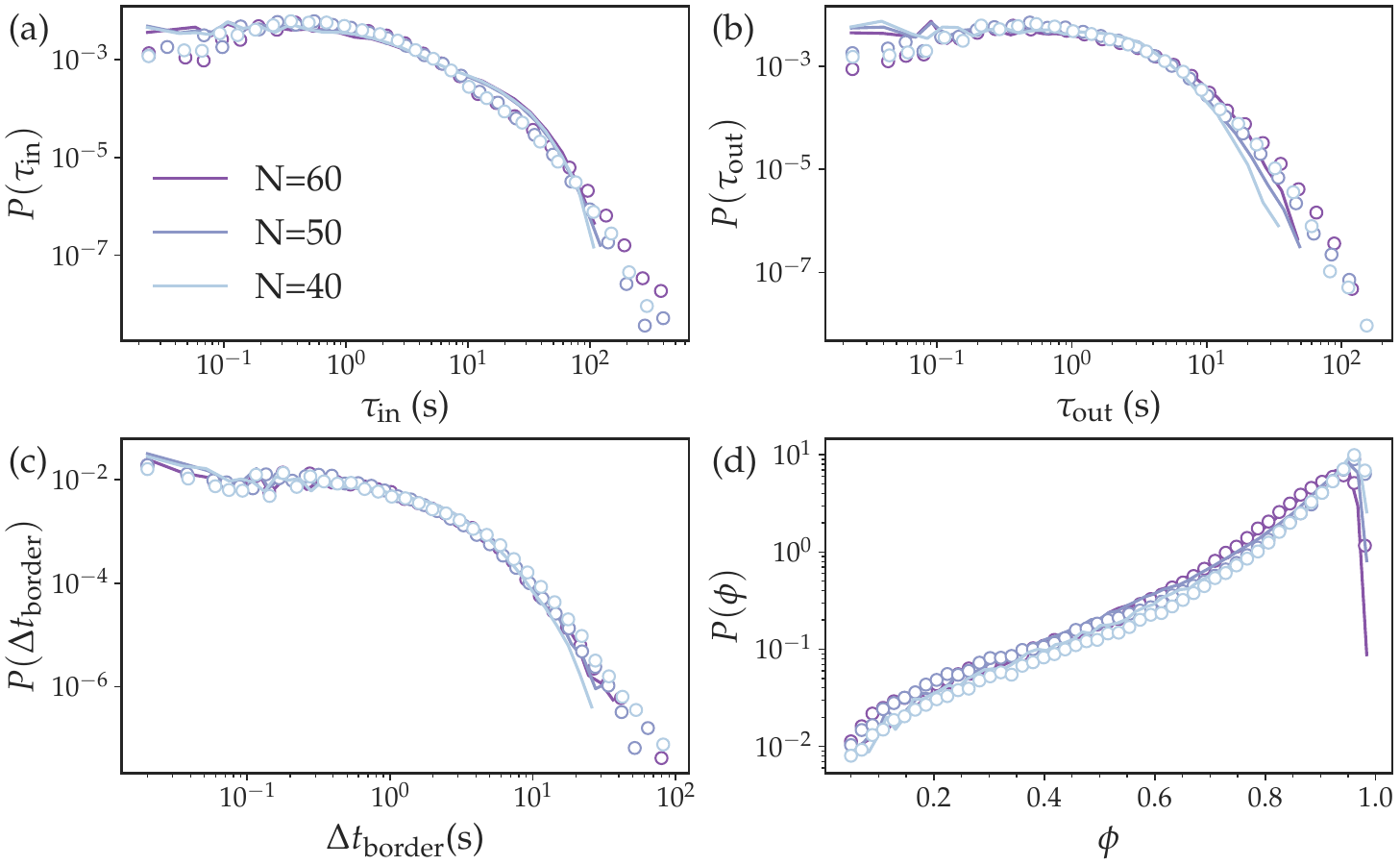}
    \caption{(a) PDFs of residence time $\tau_{\mathrm{in}}$ inside $R_g$ and (b) excursion time $\tau_{\mathrm{out}}$ outside $R_g$: experimental data (solid lines) and model simulations (dashed lines). Experimental results are aggregated over three realizations per group size. (c) PDF of peripheral residence time on the convex hull: experimental points and simulation curves. (d) PDF of polarization: experimental points compared with simulation curves. \label{fig:synth_properties}}
\end{figure}

Similarly, the excursion time outside the group, $\tau_{\mathrm{out}}$, is defined analogously as the duration a fish remains in the region $r > R_g$ after leaving. Figures~\ref{fig:synth_properties}(a) and (b) present the empirical PDFs of $\tau_{\mathrm{in}}$ and $\tau_{\mathrm{out}}$, respectively, obtained by pooling data from three experimental realizations and from matched simulations across all group sizes.
The resulting distributions are remarkably consistent across all group sizes. The model shows good agreement with experimental data for short and intermediate residence/excursion times but underestimates the probability of long excursions, i.e., it fails to reproduce the heavy tails of the empirical PDFs. Thus, while the model captures typical entering/exiting dynamics, it does not fully account for rare, large excursions away from the CM.

We examine peripheral residence following~\cite{puy_perceived_2024}. A fish is classified as peripheral if it lies on the convex hull of the group at a given frame. We then measure $\Delta t_\mathrm{border}$, the time a fish remains on the periphery (the interval between entering and leaving the convex hull) for each visit. Fig.~\ref{fig:synth_properties}(c) compares the empirical and simulated PDFs of $\Delta t_\mathrm{border}$. The model reproduces the observed distribution closely, indicating that the tendency to avoid or leave the group border is captured by the inferred individual dynamics. 

Finally, beyond individual dynamics, the model also successfully captures key collective observables used to characterize fish schools. 
One such observable is polarization,
\begin{equation}
    \phi = \left| \frac{1}{N} \sum \frac{\vec{v}_i^\mathrm{lab}}{v_i^\mathrm{lab}} \right|,
\end{equation}
which quantifies alignment of individual velocity directions~\cite{vicsek_collective_2012}. 
Polarization has been widely studied in the context of collective motion to characterize the order within groups of moving animals~\cite{vicsek_novel_1995}. Fig.~\ref{fig:synth_properties}(d) shows the empirical polarization distribution averaged over experiments for each group size. The distribution exhibits a pronounced peak near $\phi \simeq 1$, indicating a strong tendency toward directional order in black neon tetras, but also a substantial probability of low-$\phi$ states, evidenced by an exponential tail at small $\phi$. This distribution demonstrates a notable flexibility to adopt disordered configurations. Our model reproduces this bimodal tendency—high alignment with intermittent disordered configurations—yielding a polarization distribution that closely matches the experimental data.

\section*{Discussion}

In this study we introduced a stochastic framework that reproduces key statistical and dynamical properties of fish schools observed in a laboratory tank. The approach decomposes motion into two coupled layers: the school’s center-of-mass dynamics and the relative motion of individuals in the CM frame. The central hypothesis is that, in the CM reference frame, individual fish can be approximated as effectively independent particles subject to a mean-field confining potential. This effective potential encodes the net outcome of attractive and alignment tendencies, while additional empirically inferred drift and state-dependent (multiplicative) noise terms capture residual interaction effects, nonreciprocal social influences~\cite{puySelectiveSocialInteractions2024}, hydrodynamic contributions, and higher‑order correlations.
The resulting model is intentionally parsimonious yet rich enough to reproduce single‑animal statistics (speed and turning‑rate distributions, burst‑and‑coast patterns) and collective observables (polarization, spatial cohesion, residence and excursion times, MSD oscillations). Because all functional forms and coefficients are obtained from trajectory data, the framework is predictive without ad hoc tuning and provides an explicit mapping from empirical measurements to stochastic dynamical laws grounded in statistical‑physics language.

At the same time, the mean‑field character of the model implies inherent limitations. By construction it does not resolve explicit pairwise interaction kernels or short‑range neighbor structure, so it cannot capture certain fine‑scale social metrics (e.g., anisotropic neighbor maps, explicit alignment rules inferred at the pair level)~\cite{katzInferringStructureDynamics2011}. The underestimation of long excursion tails in residence/excursion-time distributions suggests that rare, large‑deviation events may require explicit modeling of correlated bursts not captured by the mean‑field potentials and inferred noise alone.

The framework suggests several concrete directions for future work. Incorporating measured heterogeneity (size, metabolic state, behavioral types) and extending the analysis to fully three‑dimensional trajectories would test the generality of the inferred laws. Controlled perturbation experiments (e.g., localized stimuli, altered confinement geometry) could probe model transferability and reveal the mechanisms underlying extreme excursions and synchronization phenomena. Finally, applying the same inference pipeline to other species and artificial swarms would clarify which features of the effective potentials and multiplicative noise are universal and which are species‑ or context‑specific.

Overall, by closing the loop between empirical trajectories, nonparametric inference, and stochastic simulation, our approach offers a principled route to translate noisy individual behavior into predictive collective dynamics. It thus contributes a compact, interpretable toolkit for studying self‑organization in biological and engineered collectives and for guiding the design and control of distributed multi‑agent systems.

\subsection*{Experimental data collection}
\label{app:exper-data-coll}

Experiments were conducted at the Scientific and Technological Centers UB (CCiTUB), University of Barcelona (Spain), and were reviewed and approved by the Ethics Committee of the University of Barcelona (project number 119/18). We used schooling black neon tetra (Hyphessobrycon herbertaxelrodi), a small freshwater species (mean body length  $\approx 2.5$ cm) that forms coherent, highly polarized, effectively planar schools~\cite{gimenoDifferencesShoalingBehavior2016}. The experiments involved schools of $N = 40, 50$ and $60$ individuals swimming freely in a experimental tank of dimensions $100 \times 100 \times 40$~cm, with a water column height of $7$~cm.  Due to this species’ tendency to swim close to the water surface, the fish movement is effectively two-dimensional.

Recordings were obtained with a GoPro Hero 11 Black at $50$ fps and a resolution of $5312\times 2988$ pixels, mounted $1$~m directly above the tank. The tank side measured $L = 2730$ pixels in the videos and was used to convert pixel coordinates to physical units. For each group size we acquired three independent $60$‑minute recordings ($180,000$ frames per recording).

Individual fish trajectories were extracted by digitizing the videos using the open-source software \texttt{idtracker.ai}~\cite{romero-ferreroIdtrackerAiTracking2019}. Occlusions causing invalid position data were corrected with the Validator tool included in version 5 of \texttt{idtracker.ai}. To reduce measurement noise, raw trajectories were smoothed with a Gaussian filter~\cite{nixonFeatureExtractionImage2010} ($\sigma=5$ frames, truncated at $5\sigma$), implemented using the \texttt{scipy.ndimage.gaussian\_filter1d} function from the \texttt{SciPy} library~\cite{virtanenSciPyFundamentalAlgorithms2020}. Individual velocities and accelerations were obtained by differentiating the smoothed trajectories using the first and second derivatives of the Gaussian kernel, respectively.

\section*{Acknowledgements}
We acknowledge financial support from projects PID2022-137505NB-C21 and PID2022-137505NB-C22, funded by MICIU/AEI/10.13039/501100011033, and by the European Regional Development Fund (ERDF): "A way of making Europe".


\end{document}